\documentclass{llncs}
\usepackage{graphicx}
\usepackage{amsmath}
\usepackage{amssymb}
\usepackage{enumerate}
\usepackage{varioref}
\usepackage{charter,eulervm}

\title{\sc {A linear-time algorithm for the strong 
chromatic index of Halin graphs}}

\author{
 Ton~Kloks\inst{1}%
\thanks{This author is supported by the
National Science Council of Taiwan, under grant
NSC~99--2218--E--007--016.}
\and
 Yue-Li~Wang\inst{2}
}
\institute{
 Department of Computer Science\\
 National Tsing Hua University,
 No.~101, Sec.~2, Kuang Fu Rd., Hsinchu, Taiwan\\
 {\tt kloks@cs.nthu.edu.tw} 
\and 
 Department of Information Management\\
 National Taiwan University of Science and Technology\\
 No.~43, Sec.~4, Keelung Rd., Taipei, 106, Taiwan\\
 {\tt ylwang@cs.ntust.edu.tw}
}

\begin{document}

\maketitle

\begin{abstract}
We show that there exists a linear-time algorithm that computes the 
strong chromatic index of Halin graphs.
\end{abstract}

\section{Introduction}

\begin{definition}
Let $G=(V,E)$ be a graph. A {\em strong edge coloring\/} of $G$ is a 
proper edge coloring such that no edge is adjacent to two edges of the 
same color. 
\end{definition}

Equivalently, a strong edge coloring of $G$ is a vertex coloring 
of $L(G)^2$, the square of the linegraph of $G$. 
The strong chromatic index of $G$ is the minimal integer $k$ such 
that $G$ has a strong edge coloring with $k$ colors. We denote the 
strong chromatic index of $G$ by $s\chi^{\prime}(G)$. 

Recently it was shown that the strong chromatic index 
is bounded by 
\[(2-\epsilon)\Delta^2\] for some $\epsilon >0$, where 
$\Delta$ is the maximal degree of the graph~\cite{kn:molloy}.%
\footnote{In their paper Molloy and Reed state that $\epsilon \geq 0.002$ 
when $\Delta$ is sufficiently large.}  
Earlier, Andersen showed that the strong chromatic index of a cubic graph 
is at most ten~\cite{kn:andersen}. 

Let $\mathcal{G}$ be the class of chordal graphs or the class of 
cocomparability graphs. If $G \in \mathcal{G}$ then also 
$L(G)^2 \in \mathcal{G}$ and it follows that the strong chromatic index 
can be computed in polynomial time for these classes. 
Also for graphs of bounded treewidth 
there exists a polynomial time algorithm 
that computes the strong chromatic 
index~\cite{kn:salavatipour}.%
\footnote{This algorithm checks in $O(n(s+1)^t)$ time whether a 
partial $k$-tree has a strong edge coloring that uses at most $s$ colors. 
Here, the exponent $t=2^{4(k+1)+1}$.} 
 
\begin{definition}
Let $T$ be a tree without vertices of degree two. Consider a plane embedding 
of $T$ and connect the leaves of $T$ by a cycle that crosses no edges of $T$. 
A graph that is constructed in this way is called a {\em Halin graph\/}. 
\end{definition}

Halin graphs have treewidth at most three. Furthermore, if $G$ 
is a Halin graph of bounded degree, then also $L(G)^2$ has 
bounded treewidth 
and thus the strong chromatic index of $G$ can be computed in 
linear time. Recently, Ko-Wei Lih, {\em et al.\/}, 
proved that a cubic Halin graph 
other than one of the two `necklaces' $Ne_2$ (the complement of 
$C_6$) and $Ne_4$, has strong 
chromatic index at most 7. The two exceptions have 
strong chromatic index 9 and 8, respectively. If $T$ is the underlying 
tree of the Halin graph, and if $G \neq Ne_2$ and $G$ is 
not a wheel $W_n$ 
with $n \neq 0 \bmod{3}$, then Ping-Ying Tsai, {\em et al.\/}, 
show that the strong chromatic index is bounded by 
$s\chi^{\prime}(T)+3$. 
(See~\cite{kn:shiu2,kn:shiu} for earlier results that  
appeared in regular papers.%
\footnote{The results of Ko-Wei Lih and Ping-Ying Tsai, {\em et al.\/}, were 
presented at the Sixth Cross-Strait Conference on Graph Theory and 
Combinatorics which was held at the National Chiao Tung University 
in Taiwan in 2011.})  
 
\medskip
If $G$ is a Halin graph then $L(G)^2$ 
has bounded rankwidth. 
In~\cite{kn:ganian} it is shown that there exists a 
polynomial algorithm that computes the chromatic number 
of graphs with bounded rankwidth, thus the strong chromatic 
index of Halin graphs can be computed in polynomial time. 
In passing, let us mention the following result. 
A class of graphs $\mathcal{G}$ is 
$\chi$-bounded if there exists a function $f$ such that 
$\chi(G) \leq f(\omega(G))$ for $G \in \mathcal{G}$. Here 
$\chi(G)$ is the chromatic number of $G$ and $\omega(G)$ is the 
clique number of $G$.  
Recently, Dvo\v{r}\'ak and Kr\'al showed that for every $k$, 
the class of graphs with rankwidth at most $k$ 
is $\chi$-bounded~\cite{kn:dvorak}. 
Obviously, the graphs $L(G)^2$ have a uniform $\chi$-bound for 
graphs $G$ in the class of Halin graphs. 

In this note we show that there exists a linear-time algorithm 
that computes the strong chromatic index of Halin graphs. 
 
\section{The strong chromatic index of Halin graphs}

The following lemma is easy to check. 

\begin{lemma}[Ping-Ying Tsai]
\label{basics}
Let $C_n$ be the cycle with $n$ vertices and let $W_n$ be the wheel 
with $n$ vertices in the cycle. 
Then 
\[s\chi^{\prime}(C_n) = \begin{cases} 
3 & \quad \text{if $n= 0 \bmod{3}$} \\
5 & \quad \text{if $n=5$} \\
4 & \quad \text{otherwise} 
\end{cases}
\quad  
s\chi^{\prime}(W_n)= \begin{cases}
n+3 & \quad \text{if $n=0 \bmod{3}$}\\
n+5 & \quad \text{if $n=5$}\\
n+4 & \quad \text{otherwise.}
\end{cases}\]
\end{lemma}

A double wheel is a Halin graph in which the tree 
$T$ has exactly two vertices 
that are not leaves. 

\begin{lemma}[Ping-Ying Tsai]
\label{basic2}
Let $W$ be a double wheel where $x$ and $y$ are 
the vertices of $T$ that are not leaves. 
Then $s\chi^{\prime}(T)=d(x)+d(y)-1$ where $d(x)$ and 
$d(y)$ are the degrees of $x$ and $y$. Furthermore, 
\[s\chi^{\prime}(W)= 
\begin{cases}
s\chi^{\prime}(T)+4=9 & \quad \text{if $d(x)=d(y)=3$, {\em i.e.\/}, if 
$W=\Bar{C_6}$} \\
s\chi^{\prime}(T)+2=d(y)+4 & \quad \text{if $d(y) > d(x)=3$} \\
s\chi^{\prime}(T)+1=d(x)+d(y) & \quad \text{if $d(y) \geq d(x) >3$.}
\end{cases}\]
\end{lemma}
  
Let $G$ be a Halin graph with tree $T$ and cycle $C$. 
Then obviously,  
\begin{equation}
\label{eq1}
s\chi^{\prime}(G) \leq s\chi^{\prime}(T) + s\chi^{\prime}(C).
\end{equation}
The linegraph of a tree is a claw-free blockgraph. Since a sun 
$S_r$ with $r > 3$ has a claw,  
$L(T)$ has no induced sun $S_r$ with $r > 3$. It follows 
that $L(T)^2$ is a chordal graph~\cite{kn:laskar} (see 
also~\cite{kn:cameron}; in this paper Cameron proves 
that $L(G)^2$ is chordal for any chordal graph $G$). Notice that 
\begin{equation}
\label{bound}
s\chi^{\prime}(T)=\chi(L(T)^2)=\omega(L(T)^2) \leq  2\Delta(G)-1 
\quad\Rightarrow\quad 
s\chi^{\prime}(G) \leq 2\Delta(G)+4.
\end{equation}

\subsection{Cubic Halin graphs}

In this subsection we outline a simple linear-time algorithm for 
the cubic Halin graphs. 

\begin{theorem}
\label{cubic case}
There exists a linear-time algorithm that computes the 
strong chromatic index of cubic Halin graphs.
\end{theorem}
\begin{proof}
Let $G$ be a cubic Halin graph with plane tree $T$ and cycle $C$. 
Let $k$ be a natural number. We describe a linear-time 
algorithm that 
checks if $G$ has a strong edge coloring with at most $k$ colors. 
By Equation (\ref{bound}) 
we may assume that $k$ is at most 10. Thus the 
correctness of this algorithm proves the theorem. 
 
\medskip
Root the tree $T$ at an arbitrary leaf $r$ of $T$.  
Consider a vertex $x$ in $T$. There is a unique path $P$ 
in $T$ from $r$ to $x$ in $T$. Define the subtree $T_x$ at $x$  
as the maximal connected subtree of $T$ that does not contain 
an edge of $P$. If $x=r$ then $T_x=T$.   

Let $H(x)$ be the subgraph of $G$ induced by the vertices of 
$T_x$. 
Notice that, 
if $x \neq r$ then the edges of $H(x)$ that are 
not in $T$ form a path $Q(x)$ of edges in $C$.   

\medskip
For $x \neq r$ define the boundary $B(x)$ 
of $H(x)$ as the following set 
of edges. 
\begin{enumerate}[\rm (a)]
\item 
\label{i1}
The unique edge of $P$ that 
is incident with $x$. 
\item 
\label{i2}
The two edges of $C$ that connect 
the path $Q(x)$ of $C$ with the rest of $C$.   
\item Consider the endpoints of the 
edges mentioned in (\ref{i1}) and (\ref{i2}) 
that are in $T_x$. 
Add the remaining two edges that are incident with each of 
these endpoints to $B(x)$. 
\end{enumerate}
Thus the boundary $B(x)$ consists of at most 9 edges. 
The following claim is easy to check. It proves 
the correctness of the algorithm described below. 
Let $e$ be an edge of $H(x)$. Let $f$ be an edge 
of $G$ that is not an edge of $H(x)$. If $e$ and $f$ are 
at distance at most 1 in $G$ then $e$ or $f$ is in 
$B(x)$.\footnote{Two edges in $G$ are at distance at most one 
if the subgraph induced by their endpoints is either $P_3$, or 
$K_3$ or $P_4$. We assume that it can be checked in constant time 
if two edges $e$ and $f$ are at distance at most one. This can be 
achieved by a suitable data structure.}
  
\medskip
Consider all possible colorings of the edges in $B(x)$. 
Since $B(x)$ contains at most 9 edges and since there are at 
most $k$ different colors for each edge, there are at most 
\[k^9 \leq 10^9\]
different colorings of the edges in $B(x)$.       

The algorithm now fills a table which gives a boolean 
value for each coloring of the boundary $B(x)$. This boolean value is 
{\tt TRUE} if and only if  
the coloring of the edges in $B(x)$ extends to an  
edge coloring of the union of the sets of edges in $B(x)$ 
and in $H(x)$ with at most $k$ colors, such that any pair of 
edges in this set that are 
at distance at most one in $G$, have different colors. 
These boolean values are computed as follows. We prove the correctness 
by induction on the size of the subtree at $x$. 

\medskip
First consider the case where the subtree at $x$ consists of the 
single vertex $x$. Then $x \neq r$ and $x$ is a leaf of $T$. 
In this case  
$B(x)$ consists 
of three edges, namely the three edges that are incident with $x$. 
These are two edges of $C$ and one edge of $T$. 
If the colors of these three edges in 
$B$  
are different 
then the boolean value is set to 
{\tt TRUE}. Otherwise it is set to {\tt FALSE}. 
Obviously, this is a correct assignment. 

\medskip
Next consider the case where $x$ is an internal vertex of $T$. 
Then $x$ has two children in the subtree at $x$. 
Let $y$ and $z$ be the two children and consider the two 
subtrees rooted at $y$ and $z$. 

The algorithm that computes the tables for each vertex $x$ 
processes the subtrees in order of increasing number of vertices.   
(Thus the roots of the subtrees are visited in postorder). 
We now assume that the tables at $y$ and $z$ are 
computed correctly and show how the  
table for $x$ is computed correctly and in constant time. 
That is, we prove that the algorithm described below 
computes the table at $x$ such that it contains a 
coloring of $B(x)$ with a value {\tt TRUE} if and only 
if there 
exists an extension of this coloring to the 
edges of $H(x)$ and $B(x)$ such that any two 
different edges $e$ and $f$ 
at distance at most one in $G$, each one in 
$H(x)$ or in $B(x)$, have different colors. 
   
\medskip
Consider a coloring of the edges in the 
boundary $B(x)$. 
The boolean value in the table of $x$ 
for this coloring is computed as follows. 
Notice that 
\begin{enumerate}[\rm (i)]
\item $B(y) \cap B(z)$ consists of one edge and this 
edge is not in $B(x)$, and 
\item $B(x) \cap B(y)$ consists 
of at most four edges, namely the edge $(x,y)$ and the three 
edges of $B(y)$ that are incident with one vertex of 
$C \cap H(y)$.   
Likewise, $B(x) \cap B(z)$ consists of at most four edges. 
\end{enumerate} 
The algorithm varies the possible colorings of 
the edge in $B(y) \cap B(z)$. 
Colorings of $B(x)$, $B(y)$ and $B(z)$ are 
consistent if the intersections are the same color and the pairs 
of edges in 
\[B(x) \cup B(y) \cup B(z)\] 
that are at distance at most one in $G$ have different colors.   
A coloring of $B(x)$ is assigned the value {\tt TRUE} 
if there exist colorings of $B(y)$ and $B(z)$ such that the 
three colorings are consistent and $B(y)$ and $B(z)$ are assigned the 
value {\tt TRUE} in the tables at $y$ and at $z$ respectively. 
Notice that the table at $x$ is built in constant time. 

Consider a coloring of $B(x)$ that is assigned the value {\tt TRUE}. 
Consider colorings of the edges of $B(y)$ and $B(z)$ that are 
consistent with $B(x)$ and that are assigned the value 
{\tt TRUE} in the tables at $y$ and $z$. By induction, there 
exist extensions of the colorings of $B(y)$ and $B(z)$ 
to the edges of $H(y)$ and $H(z)$. The union of these 
extensions provides a $k$-coloring of the edges in $H(x)$. 

Consider two edges $e$ and $f$ in $B(x) \cup B(y) \cup B(z)$. 
If their distance 
is at most one then they have different colors since 
the coloring of $B(x) \cup B(y) \cup B(z)$ is 
consistent. Let $e$ and $f$ be a pair of edges in $H(x)$. 
If they are both in $H(y)$ or both in $H(z)$ then they have 
different colors.  
Assume that $e$ is in $H(y)$ and assume that $f$ is not in 
$H(y)$. If $e$ and $f$ are at distance at most one, 
then $e$ or $f$ is in $B(y)$. If they are both in $B(y)$, 
then they have different colors, due to the consistency. 
Otherwise, by the induction hypothesis, they have different colors. 
This proves the claim on the correctness. 
    
\medskip
Finally, consider the table for the vertex $x$ which is 
the unique neighbor of $r$ in $T$. 
By the induction hypothesis, 
and the fact that every edge in $G$ is either in 
$B(x)$ or in $H(x)$,   
$G$ has a strong edge coloring with at most 
$k$ colors if and only if the table at $x$ contains a coloring 
of $B(x)$ with three different colors for which the boolean is 
set to {\tt TRUE}. 

This proves the theorem. 
\qed\end{proof}

\begin{remark}
The involved constants in this algorithm are improved considerably 
by the recent results of Ko-Wei Lih, Ping-Ying Tsai, {\em et al.\/}. 
\end{remark}
  
\subsection{Halin graphs of general degree}

\begin{theorem}
\label{general Halin graphs}
There exists a linear-time algorithm that computes the strong 
chromatic index of Halin graphs.
\end{theorem}
\begin{proof}
The algorithm is similar to the algorithm for 
the cubic case.   

\medskip
Let $G$ be a Halin graph, let $T$ be the underlying plane tree,  
and let $C$ be the cycle that connects the leaves of $T$. 
Since $L(T)^2$ is chordal the chromatic number of $L(T)^2$ 
is equal to the 
clique number of $L(T)^2$, which is 
\[s\chi^{\prime}(T)= \max \;\{\;d(u)+d(v)-1\;|\; (u,v) \in E(T)\;\},\]
where $d(u)$ is the degree of $u$ in the tree $T$. 
By Formula~(\ref{eq1}) and Lemma~\ref{basics} 
the strong chromatic index of $G$ is one of 
the six possible values%
\footnote{Actually, according to the recent results of 
Ping-Ying Tsai, {\em et al.\/}, 
the strong chromatic index of $G$ is at most $s\chi^{\prime}(T)+3$ 
except when $G$ is a wheel or $\Bar{C_6}$.}
\[s\chi^{\prime}(T), s\chi^{\prime}(T)+1, \ldots, s\chi^{\prime}(T)+5.\]

\medskip
Root the tree at some leaf $r$ and consider a subtree $T_x$ at a 
node $x$ of $T$. 
Let $H(x)$ be the subgraph of $G$ induced by the vertices of $T_x$.  
Let $y$ and $z$ be the two boundary vertices of $H(x)$ in $C$. 

\medskip
We distinguish the following six types of edges corresponding 
to $H(x)$. 
\begin{enumerate}[\rm 1.]
\item The set of edges in $T_x$ that are adjacent to $x$. 
\item The edge that connects $x$ to its parent in $T$. 
\item The edge that connects $y$ to its neighbor in $C$ that is 
not in $T_x$. 
\item The set of edges in $H(x)$ that have endpoint $y$. 
\item The edge that connects $z$ to its neighbor in $C$ that is 
not in $T_x$. 
\item The set of edges in $H(x)$ that have endpoint $z$. 
\end{enumerate}

Notice that the set of edges of every type has bounded 
cardinality, except the first type. 

\medskip
Consider a $0/1$-matrix $M$ with rows indexed by the six 
types of edges and columns indexed by the colors. 
A matrix entry $M_{ij}$ is 1 if there is an edge of the row-type 
$i$ that is colored with the color $j$ and otherwise this 
entry is 0. Since $M$ has only 6 rows, the rank over 
$GF[2]$ of $M$ is at most 6.  

\medskip 
Two colorings are equivalent if there is a 
permutation of the colors that maps one coloring to the other one. 
Let $S \subseteq \{1,\ldots,6\}$ and let $W(S)$ be the 
set of colors that are used by 
edges of type $i$ for all $i \in S$.  
A class of equivalent colorings is fixed by the set 
of cardinalities  
\[\{\; |W(S)| \;|\; S \subseteq \{1,\ldots,6\} \;\}.\]

\medskip
We claim that the number of equivalence classes is 
constant. The number of ones in the row of the first type 
is the degree of $x$ in $H(x)$. Every other row has at most 
3 ones. This proves the claim. 

\medskip 
Consider the union of two subtrees, say at $x$ and $x^{\prime}$. 
The algorithm considers all equivalence classes of colorings of the 
union, and checks, by table look-up, whether it decomposes into 
valid colorings of $H(x)$ and $H(x^{\prime})$. An easy way to do this 
is as follows. First double the number of types, by distinguishing 
the edges of $H(x)$ and $H(x^{\prime})$. Then enumerate 
all equivalence classes of colorings. Each equivalence class 
is fixed by a sequence of $2^{12}$ numbers, as above. 
By table look-up,  
check if an equivalence class restricts to a valid coloring 
for each of $H(x)$ and $H(x^{\prime})$.  
Since this takes constant time, the algorithm runs in linear time. 

This proves the theorem. 
\qed\end{proof}


\begin{thebibliography}{99}

\bibitem{kn:andersen}Andersen,~L., 
The strong chromatic index of a cubic graph is at most 10, 
{\em Discrete Mathematics\/} {\bf 108} (1992), pp.~231--252. 

\bibitem{kn:cameron}Cameron,~K., 
Induced matchings, 
{\em Discrete Applied Mathematics\/} {\bf 24} (1989), pp.~97--102. 

\bibitem{kn:dvorak}Dvo\v{r}\'ak,~Z. and D.~Kr\'al, 
Classes of graphs with small rank decompositions are $\chi$-bounded. 
Manuscript on ArXiv: 1107.2161.v1, 2011. 

\bibitem{kn:faudree}Faudree,~R., A.~Gy\'arf\'as, R.~Schelp, Z.~Tuza, 
The strong chromatic index of graphs, 
{\em Ars Combinatorica\/} {\bf 29B} (1991), pp.~205--211. 

\bibitem{kn:ganian}Ganian,~R. and P.~Hlin\v{e}n\'y, 
Better polynomial algorithms on graphs of bounded rankwidth, 
{\em Proceedings IWOCA~09\/}, LNCS~5874 (2009), pp.~266--277. 

\bibitem{kn:golumbic}Golumbic,~M. and M.~Lewenstein, 
New results on induced matchings, 
{\em Discrete Applied Mathematics\/} {\bf 101} (2000), pp.~157--165. 
 
\bibitem{kn:halin}Halin,~R., 
Studies on minimally $n$-connected graphs, 
in (D.~Welsh ed.) 
{\em Combinatorial mathematics and its applications\/}, 
Academic Press (1971), pp.~129--136. 

\bibitem{kn:ka}Ka,~T., 
{\em The strong chromatic index of cubic Halin graphs\/}. 
M.Phil. Thesis, Hong Kong Baptist University, 2003. 

\bibitem{kn:laskar}Laskar,~R. and D.~Shier, 
On powers and centers of chordal graphs, 
{\em Discrete Applied Mathematics\/} {\bf 6} (1983), 
pp.~139--147. 
 
\bibitem{kn:mahdian}Mahdian,~M., 
{\em The strong chromatic index of graphs\/}. 
M.Sc. Thesis, Department of Computer Science, 
University of Toronto, 2000. 

\bibitem{kn:mahdian2}Mahdian,~M., 
On the computational complexity of strong edge coloring,
{\em Discrete Applied Mathematics\/} {\bf 118} (2002), pp.~239--248. 

\bibitem{kn:molloy}Molloy,~M. and B.~Reed, 
A bound on the strong chromatic index of a graph, 
{\em Journal of Combinatorial Theory, Series B\/} {\bf 69} (1997), 
pp.~103--109. 

\bibitem{kn:salavatipour}Salavatipour,~M., 
A polynomial algorithm for strong edge coloring 
of partial $k$-trees, 
{\em Discrete Applied Mathematics\/} {\bf 143} (2004), pp.~285--291. 

\bibitem{kn:shiu2}Shiu,~W., P.~Lam and W.~Tam, 
On strong chromatic index of Halin graphs, 
{\em Journal of Combinatorial Mathematics and Combinatorial Computing\/} 
{\bf 57} (2006), pp.~211-222. 

\bibitem{kn:shiu}Shiu,~W. and W.~Tam, 
The strong chromatic index of complete cubic Halin graphs, 
{\em Applied Mathematics Letters\/} {\bf 22} (2009), pp.~754--758. 

\bibitem{kn:wu}Wu,~J. and W.~Lin, 
The strong chromatic index of a class of graphs, 
{\em Discrete Mathematics\/} {\bf 308} (2008), pp.~6254--6261. 

\end{thebibliography}
\end{document}